\documentclass[11pt]{amsart}
\allowdisplaybreaks[1]

\usepackage{amssymb}
\usepackage{amsmath}
\usepackage{amscd}
\usepackage{epstopdf}
\usepackage{graphicx}

\theoremstyle{plain}
\newtheorem{theorem}{Theorem}
\newtheorem{lemma}{Lemma}
\newtheorem{prop}{Proposition}

\theoremstyle{definition}

\newcommand{\ts}{\hspace{0.5pt}}

\newcommand{\CC}{\mathbb{C}\ts}
\newcommand{\RR}{\mathbb{R}\ts}

\newcommand{\ZZ}{\mathbb{Z}}
\newcommand{\NN}{\mathbb{N}}

\newcommand{\TT}{\mathbb T}

\newcommand{\Rd}{{\mathbb R}^d}

\newcommand{\chF}{{\boldsymbol 1}}

\newcommand{\vL}{\varLambda}
\newcommand{\gL}{\Lambda}

\newcommand{\gO}{\Omega}

\newcommand{\cL}{\mathcal{L}}
\newcommand{\CalU}{\mathcal{U}}
\newcommand{\CalD}{\mathcal{D}}

\newcommand{\CalF}{\mathcal{F}}

\newcommand{\CalS}{\mathcal{S}}

\newcommand{\dens}{\mathrm{dens}}

\newcommand{\card}{{\mathrm{card}}}

\newcommand{\Vol}{\ell}
\newcommand{\vol}{\mathrm{vol}}

\newcommand{\frq}{{\rm freq}}

\newcommand{\bU}{{\bf U}}

\begin{document}
\title[Model sets and their point correlations]
{How model sets can be determined by their two-point and three-point correlations}

\author{Xinghua Deng and Robert V.\ Moody}
\address{Department of Mathematics and Statistics,
University of Victoria, \newline \hspace*{12pt}Victoria, BC, V8W3P4,
Canada} \email{rmoody@uvic.ca, x.deng@open.ac.uk}
\date{\today}

\begin{abstract}
We show that real model sets with real internal spaces are determined, up
to translation and changes of density $0$, by their $2$- and $3$-point
correlations. We also show that there exist pairs of real (even $1D$) aperiodic model
sets with internal spaces that are products of real spaces and finite cyclic groups
whose $2$- and $3$-point correlations are identical but which are not related by
either translation or inversion of their windows. All these examples are pure
point diffractive.

Placed in the context of ergodic uniformly discrete point processes, the result is that
real point processes of model sets based on real internal windows are determined
by their second and third moments.
\end{abstract}
\maketitle

\section{Introduction}

An enduring problem of crystallography is the
inference of the internal structure of a crystal from physically measurable
quantities, notably diffraction. Perfect knowledge of the diffraction is equivalent to perfect knowledge of the $2$-point correlation of the structure. However, even
perfect knowledge of the diffraction, or $2$-point correlation, is insufficient
to pin down the structure of a crystal, with counterexamples going back to L.~Patterson \cite{Patterson44} (see \cite{GM} for a good source of information on this subject).

Quasicrystals present the same problem,
but are even more difficult. Based on the theory of the covariogram, Baake and Grimm \cite{BG, BG2} have given examples of model sets
(or cut and project sets, as they are often called), which are intrinsically different but have the same diffraction.

One can ask whether knowledge of additional higher point correlations, notably the
$3$-point correlation measure, could
provide sufficient information determine the structure. But again, even for periodic structures, there are counterexamples \cite{GM}. The main result of this paper
(Thm.~\ref{main})
however, is a positive result: in the context of regular model sets with
real internal spaces, the $2$- and $3$-point correlations do determine the model set (up to translation and to modification by sets of density $0$).

The possibility that in pure point diffractive sets (of which regular model
are good examples) the $2$- and $3$-point correlations would suffice to know
all the higher correlations was suggested by D.~Mermin
in a very interesting paper \cite{Mermin} on a new approach to handling
symmetry for quasicrystals. His ideas are based on the Landau
approach to second order phase transitions and involve rather informal
manipulation of quantities which, as the author recognizes, cannot be mathematically justified as given.
To quote from that paper: \emph{ ... this informal Fourier space
argument that the identity of all second and third order correlations
implies the identity of all higher correlations is disarmingly trivial. I would
very much like to learn of a comparably simple informal argument
or an instructive counterexample in position space.}

In spite of Thm.~\ref{main},  Mermin's suggestion is not correct in general. This can already be seen in the periodic case from the results of \cite{GM}, which we illustrate here in \S\ref{PE} in the form of periodic model sets.
In \S\ref{Aperiodic} we offer an example of a pair of aperiodic model sets, based on an internal space which is the direct product of a real line and a finite cyclic group, for
which the $2$- and $3$-point correlations are identical but the point sets
themselves cannot be transformed into each other either by translation or inversion of their
windows.

The situation with more general internal spaces is, no doubt,
difficult.  Even the simple case of a product of a real space and a
finite group alluded to above is quite complicated. We touch on it
here in \S\ref{CE}.  Although, as we
have just pointed out, counterexamples can occur in this case, in other
instances we can obtain a positive result. For instance, it is easy to see that the
vertices of rhombic Penrose tilings, which are model sets based on $\RR^2 \times
\ZZ/5\ZZ$, are determined by their $2$-and $3$-point correlations.

Aperiodic sets are often studied in the context of dynamical systems and/or stochastic point processes. This approach was pioneered in \cite{Radin} and has been used extensively both in mathematical and physical models, \cite{Bellissard,
Gouere, DM}. This applies particularly to regular model sets (see \S\ref{ms} for definitions) where things can be stated much more precisely, \cite{Hof, Schlottmann, BLM}. Instead of a single model set one considers its hull, namely the set of all
uniformly discrete points sets that are in the closure of its translation orbit (see
\S\ref{cpp} for more details). This hull is then uniquely ergodic, and we may view
it as describing a uniformly discrete ergodic point process.

Any such uniformly discrete point process (with a common lower bound on the distance between closest points)  is characterized by knowledge of its entire set of moments (second, third, etc.)\cite{DM}.
Knowledge of the $k$th moment is equivalent to knowledge of the $k$-point correlation.
Thus our result about model sets with real internal spaces can be rewritten
(Thm.~\ref{mainPointProcess}) as the statement that for them only finitely many (namely the second and third)
of these infinitely many moments are needed.

The main idea behind proving Thm.~\ref{main} is to transfer the
correlation problem to the internal space of the cut and project scheme
defining the model sets. There the correlations are directly related to
what have been called in \cite{JK} the $k$-deck functions, or the
covariogram in the case $k=2$.  These become tractable after Fourier transformation,
a fact that has been discovered several times before, see
for example \cite{GM, JK}. The main obstacle is dealing with the set $E$ of zeros (the {\em extinctions} in the diffraction) of the Fourier transform of the characteristic function of the window of the model set. Here we offer Prop.~\ref{ppcharacter} which we have not seen
explicitly stated in the literature, which allows us to proceed as long as $E$ has no interior points.

Finally in \S\ref{cpp} we point out \cite{LenzM} which offers a different approach via
spectral theory to determining uniformly discrete ergodic {\em pure point diffractive} point processes by means of their moments.  However, ultimately it too returns to similar problems about extinctions. For a short survey covering this and material on point processes, see \cite{MoodyICQ}.

\section{Model sets} \label{ms}

We work in $\RR^d$. The usual Lebesgue measure will be denoted by $\Vol$.
The open cube of side length $R$ centred at $0$ is denoted by $C_R$, so
$\ell(C_R) = R^d$.
A {\em cut and project scheme} for $\RR^d$ is a triple
$\CalS = (\RR^d, H, \cL)$ consisting of a compactly generated locally compact Abelian group $H$ and a lattice $\cL \subset \RR^d \times H$ for which the
projection mappings
$\pi_1$ and $\pi_2$ from $\RR^d \times H$ onto
$\RR^d$ and $H$ are injective and have dense image respectively:

\begin{equation} \label{cpScheme}
   \begin{array}{ccccc}
      \RR^d & \stackrel{\pi_{1}}{\longleftarrow} &
        \RR^d \times H & \stackrel{\pi_{2}}
      {\longrightarrow} & H   \\
      &&  \cup \\
      L&\stackrel{\simeq}{\longleftrightarrow} & \cL &\qquad &\qquad \,.
   \end{array}
\end{equation}

Then $L:=\pi_1 (\cL)$ is isomorphic as a group to $\cL$ (though
it is rarely a discrete subgroup of $\RR^d$) and we have the composite
mapping $(\cdot)^\star \! : \, L \longrightarrow H$ with dense image
defined by $\pi_2 \circ (\pi_1|_{\cL})^{-1}$.

The statement that $\cL$ is a lattice is equivalent to saying that it
is a discrete subgroup of $\RR^d \times H$ and that the
quotient group $\TT := (\RR^d \times H)/\cL$ is compact.
Notice that if $H= \{0\}$ then $L$ is a lattice in $\RR^d$ and we are back in the
situation of normal periodic crystallography. Thus the theory of model sets is
a generalization of the theory of periodic sets and includes them as special
cases.

We let $\theta_H$ be a Haar measure on $H$, scaled so it gives measure $1$ to any fundamental region of $\cL$ in the space $\RR^d \times H$ under the product measure
$\ell \otimes \theta_H$. This is the same as saying that the naturally induced measure
$\theta_\TT$ on $\TT$ is normalized
so that $\theta_\TT (\TT) = 1$. This normalization leads to the
uniform distribution equation \eqref{udFormula} in the form given in Thm.~\ref{uniformDistribution} below.\footnote{In particular situations, for instance if the internal space is a real space, one may prefer to normalize on the basis of what seems natural for $\theta_H$ and thereby introduce
a multiplicative factor into \eqref{udFormula}. This amounts to corresponding variations by multiplicative factors in the correlations and frequencies in which we are interested, but has no intrinsic importance to what we are discussing here.}

 For $W \subset H$,
\[ \vL(W):= \{ u\in L: u^\star \in W\} \,. \]

 A set $\gO \subset H$ is called a {\em window} if
$\Sigma^\circ \subset \gO \subset \Sigma$ for some compact
set $\Sigma \subset H$ which satisfies $\overline{\Sigma^\circ} =\Sigma$.
We shall often deal with families of windows based on a single
$\Sigma$.
A {\em model set} or {\em cut and project set} is a set of the form
\[ \vL(x,y)= \vL(x,y,\gO):= x+  \vL(-y +\gO) \,,\]
 where $\gO$ is a window and $(x,y) \in \RR^d \times H$.  The model
set is called {\em regular} if the boundary of $\gO$ (or equivalently of
$\Sigma$) has measure $0$  with respect
to the Haar measure of $H$.

It is easy to see that for a fixed
$\gO$, $\vL(x,y,\Omega)$ depends only on $\xi := (x,y) \mod \cL$, so we may write
$\vL(\xi,\Omega)$ instead.

The cut and project scheme $\CalS$ together with a compact
window $\Sigma$ is called {\em irredundant} if
the equation $t + \Sigma = \Sigma$ implies that $t=0$. When dealing with
model sets it is always possible to adjust the
cut and project scheme (by factoring out a subgroup of $H$) to get
an irredundant cut and project scheme which defines exactly the same
family  $\vL(\xi, \Omega)$, $\Sigma^\circ \subset \gO \subset \Sigma$,
though now $\Sigma$ is replaced by its image in the quotient of $H$
\cite{Schlottmann, BLM, LM}.

If $\CalS$ and $\CalS'$ are irredundant cut and project schemes for
the same model set (or ones differing on sets of density $0$) then
their internal spaces $H,H'$ are isomorphic topological groups
by an isomorphism that induces an isomorphism of the corresponding
lattices in the obvious way. Thus one may speak of {\em the} irredundant cut
and project scheme of a model set. The proof of this essentially follows
from the construction of $H$ given in \S\ref{c2ms}, see
\cite{LM,BLM}. We shall always assume that
we are in the irredundant situation.

Model sets $\Lambda$, regular or not, are Delone subsets of $\RR^d$, that is to say,
there exist positive real numbers $r,R$ so that the cubes
$C_r, C_R$ of side lengths $r$ and $R$, no matter where they
are translated to in $\RR^d$, have
at most one point of (respectively at least one point of) $\Lambda$.
Since any model set
$\gL = \vL(x,y,\gO)$ also satisfies $\gL - \gL \subset \vL(\gO-\gO)$
(sets of differences) and since $\gO-\gO$ is relatively compact, we see
that $\gL - \gL $ is also uniformly discrete, and by a similar
arguments, all finite sets of sums and differences
\[ \gL \pm \dots \pm \gL \quad (\mbox{$n$ terms})\]
are also uniformly discrete. This is the {\em Meyer} property of model sets
\cite{MoodyNato}.

Model sets are uniformly distributed point sets:

\begin{theorem} \cite{Moody} \label{uniformDistribution}
Let $\Omega \subset H$ be measurable and relatively compact.
Then, assuming the normalization of measures assumed in \S\ref{ms} .
\begin{equation}\label{udFormula}
\lim_{R\to\infty} \frac{1}{\Vol (C_R)} \card (\vL(\xi,\gO) \cap C_R)
= \theta_H(\gO)
\end{equation}
for $\xi \in \TT$, $\theta_\TT$ almost surely. If the boundary
of $\gO$ has Haar measure $0$ then the result holds for
all $\xi \in \TT$. \qed
\end{theorem}

{\bf Remarks}:   \begin{itemize}
\item[(i)] When we write
$\lim_{R\to\infty}\frac{1}{\Vol (C_R)}$ we mean that
we use a (any) discrete increasing sequence of positive real numbers $\{R_j\} \to
\infty$.

\item[(ii)]
In this paper we need to work with averaging sequences
like the one in Thm. ~\ref{uniformDistribution}.
The results here apply for any van Hove sequence $\{ A_n\}$ of
subsets of $\RR^d$ satisfying
the condition that there is a constant $C>0$ with
$\Vol(A_n - A_n) \le C \Vol(A_n) $. In the case of regular model sets, which
are the main focus of this paper,
all results quoted that depend on averaging (frequencies, correlations) are
actually independent of which averaging sequence we use. This is a consequence of unique ergodicity.

\item[(iii)] In particular, although all averaging results
in this paper are written with averages over $C_R$, the centres need
not be restricted to $0$ and cubes could be replaced by balls, etc.
\end{itemize}

\section{Correlations and diffraction}

Let $\gL = \vL(\xi, \gO)$ be a regular model set. The {\em frequency}
in $\gL$ of a set
of points $\{0, x_1, \dots, x_n\} \subset \RR^d$ is defined as
\[\frq(\{0, x_1, \dots, x_n\}) = \lim_{R\to\infty} \frac{1}{\Vol(C_R)} \card \{y \in C_R\, : \, y,  y+x_1, \dots, y+x_n  \in \gL  \} \, .\]
The frequency is the expected number of occurrences of the pattern per unit
of vollume in $\RR^d$. There is no need for the elements of $\{0, x_1, \dots, x_n\}$
to be distinct, though repetitions can clearly be deleted.

The following is a well known consequence of the uniform distribution of
model sets:

\begin{prop} \label{frequency}
Let $\Sigma \subset H$ be non-empty and compact with
$\Sigma = \overline{\Sigma^\circ}$ and
$\theta_H(\partial \Sigma) = 0$. Let $\gL = \vL(\xi, \gO)$ be a regular model set
with $\Sigma^\circ \subset \gO \subset \Sigma$.
Then the frequency of $\{0, x_1, \dots, x_n\}$ in $\gL = \vL(\xi, \gO)$
exists and
\[\frq(\{0, x_1, \dots, x_n\}) = \theta_H(\gO  \cap\bigcap_{j=1}^n (-x_j^\star + \gO) )
=  \theta_H(\Sigma \cap\bigcap_{j=1}^n (-x_j^\star + \Sigma) )\]
for all $\xi \in \TT$. In particular, the frequency of
$\{0, x_1, \dots, x_n\}$ in $\vL(\xi, \gO)$ does not depend on $\xi$ but only on
the cut and project scheme $\CalS$ and the closure $\Sigma=\overline{\gO}$ of the window.
\end{prop}

{\bf Proof:}  First assume the simple
case that $\gL = \vL(\gO)$. Then
$y, y+x_1, \dots, y+ x_n \in \gL$ iff
$y^\star, y^\star + x_1^\star, \dots, y^\star + x_n^\star \in \gO$
iff $y^\star \in W:= \gO \cap\bigcap_{j=1}^n (-x_j^\star + \gO)$. The quantity
we are looking for is then
\[\lim_{R\to\infty} \frac{1}{\Vol (C_R)} \card (\vL(W) \cap C_R)
= \theta_H(W) \,, \]
which proves the first claim in this case. In the general setting,
we are looking at $x + \vL(-y+ \gO)$, but it is clear that these
translations do not affect the outcome. Nor does replacing
$\gO$ by $\Sigma$, which only results in measure $0$ changes
to the set $W$. \qed

\smallskip

The $(n+1)$-point {\bf correlation} ($n = 1,2, \dots$) of a model set
$\gL$ (or more generally any locally finite subset of $\RR^d$) is the measure
on $(\Rd)^n$ defined by
\begin{eqnarray*}
\gamma_\gL^{(n+1)}(f)&=&\lim_{R\to\infty} \frac{1}{\Vol (C_R)}\sum_{y_1,\dots, y_n,x
\, \in \,C_R \cap \gL} T_xf(y_1, \dots,y_n)\\
&=&\lim_{R\to\infty} \frac{1}{\Vol (C_R)}\sum_{\stackrel{x\in C_R \cap \gL}{ y_1,\dots y_n \in \gL}} T_xf( y_1, \dots, y_n) \, ,
\end{eqnarray*}
for all $f\in C_c((\Rd)^n)$. Here $T_x$ is simultaneous translation of all the variables
by $x$.
The simpler second sum is a result of the van Hove property of the averaging
sequence $\{C_R\}$. Because model sets are Meyer sets, the sets of elements
$y_j-x $ which make up the values of the arguments of $f$ occuring
in the sums, lie in the {\em uniformly discrete} set $\gL - \gL$. Hence for model
sets we find that
\begin{equation}\label{eqcorrelation3}
  \gamma_\gL^{(n+1)} = \sum_{x_1,\dots,x_n\in \gL -\gL}
\frq(\{0,x_1,\dots,x_n\}) \, \delta_{(x_1,\dots,x_n)} \, .
\end{equation}

In view of Prop.~\ref{frequency}, all the correlations of model sets exist
and they depend only on the closure of the window. In other words, for a
given compact window $\Sigma$ which is the closure of its interior, all model sets $\vL(\xi, \gO)$ with $\Sigma^\circ \subset \gO \subset \Sigma$ have the same
correlations of all orders. Thus we can collect the model sets into families
$\CalF(\CalS, \Sigma)$ which all have the same correlations.

The {\em diffraction} of a point set in $\RR^d$ is, by definition, the Fourier transform
$\widehat{\gamma_\gL^{(2)}}$
of its $2$-point correlation. In the case of model sets, this can be described explicitly.
For the given cut and project scheme $\CalS$, see \eqref{cpScheme},
there is a Fourier dual $\widehat{\CalS}$ of it:

\begin{equation} \label{dualcpScheme}
   \begin{array}{ccccc}
      \widehat{\RR^d} & \stackrel{\pi_{1}}{\longleftarrow} &
        \widehat{\RR^d} \times \hat H & \stackrel{\pi_{2}}
      {\longrightarrow} & \hat H   \\
      &&  \cup \\
      L^\circ&\stackrel{\simeq}{\longleftrightarrow} & \hat \TT &\qquad &\qquad \,,
   \end{array}
\end{equation}
formed by taking the Fourier duals of the groups involved in $\CalS$
\,\cite{MoodyNato}.
All of these groups, being duals of locally compact Abelian groups, are also
locally compact and Abelian. Although $\widehat{\RR^d}$ is canonically isomorphic with $\RR^d$, sometimes, for the sake of clarity, it is convenient to make the notational distinction, as we do
here.

Here $\pi_1$ and $\pi_2$ again are projections. The dual of the compact
group $\TT$ is discrete, and $\pi_1|_{\hat{\TT}}$ is injective. The image of
$\hat \TT$ is denoted by $L^\circ$. There is again a mapping $^\star: L^\circ \longrightarrow \hat H$.

The diffraction of any regular model set defined by a window $\gO$ is pure
point \cite{Schlottmann} and
is supported on a subset of $L^\circ$ of $\widehat{\RR^d}$; it is explicitly
given by
\begin{equation} \label{diffractionEq}
\widehat{\gamma_\gL^{(2)}}(\{k\}) = | \widehat{\chF_\gO(-k^\star})|^2 \,.
\end{equation}

The set of $k \in \widehat{\RR^d}$ at which the diffraction is not zero
(and hence is actually positive) is the set of locations of the Bragg peaks  of
$\gL$ (the {\em Bragg peaks} being the combined information of the location and intensity
of the atomic part of the diffraction).

We wish to prove that the $2$- and $3$-point correlations completely
classify all such families $\CalF(\CalS, \Sigma)$; that is to say, given that a point set {\em is} a model
set in $\RR^d$, then the $2$- and $3$-point correlations determine the cut and project
scheme $\CalS$ and the compact window $\Sigma$ up to translation. We begin
with an abstract result on extending partially defined group characters.

\section{Extending partial characters}\label{EC}

Let $G$ be a locally compact Abelian group and let $U(1)$
denote the compact group which is the unit circle in $\CC$. The
dual to $G$ is the group $\hat G$ of all continuous characters, i.e.
continuous homomorphisms of $G$ into $U(1)$.

Suppose that $E$ is a
closed subset of $G$ with no interior, and $0 \notin E$. Let $D:=
G\backslash E$ and $D^{(2)}:=\{(k_1,k_2):k_1,k_2,k_1+k_2\in D\}$.

\begin{lemma}\label{lemdense}
$D^{(2)}$ is dense in $G\times G$.
\end{lemma}
{\sc Proof:} Suppose $D^{(2)}$ is not dense in $G\times G$, i.e.,
there is a non-empty open set $U\times V\subset (G\times G)\setminus D^{(2)}$,
where $U, V\subset G$ are open. Since $E$ is a closed subset of $G$
with no interior, $D\cap U, D\cap V$ are nonempty open sets. For all
$u\in D\cap U, v\in D\cap V$, we have $(u,v)\in (G\times G)\setminus
D^{(2)}$, i.e., $u+v\in E$. Thus, $D\cap U\subset -v+E$. This is
impossible since $D\cap U$ is an open set and $E$ has no interior. \qed

\begin{prop}\label{ppcharacter}
Let
\[ \varphi: D \longrightarrow U(1) \]
be a continuous mapping satisfying
\[\varphi (s+t) = \varphi(s) \varphi(t)\]
whenever $s,t,s+t \in D$. Then there is a unique character $\chi \in
\widehat G$ with $\chi|D = \varphi$.
\end{prop}

{\bf Proof:} Let $\CalU$ be the uniformity on $G$ defined by its
structure as a topological group: the basic entourages are the sets
\[\bU(V) := \{(x,y) : G \times G, x-y\in V\} \]
where $V$ runs through open neighbourhoods of $0 \in G$.
This uniformity also induces the relative uniformity on
$D$ by intersecting its entourages with $D\times D$,
and this relative uniformity induces the relative
topology on $D$ \cite{Kelly}, Ch.6.

We claim that $\varphi: D \longrightarrow U(1) $ is uniformly
continuous. We show that given any $\epsilon >0$ there is an
entourage $\bU(V(\epsilon)) \cap (D\times D)$ for which $(s,t) \in
\bU(V(\epsilon))\cap (D\times D)$ implies that $|\varphi(s)
-\varphi(t)| < \epsilon$.

In fact $V(\epsilon) := \{ s \in D: |\varphi(s) -1| < \epsilon \}$
works. This is an open subset of $D$ containing $0$ and furthermore,
$(s,t) \in \bU(V(\epsilon))\cap (D\times D) $ implies $s-t \in
V(\epsilon)$ and then $s-t \in D$ and $|\varphi(s-t) -1| <\epsilon$.
Using the basic relation satisfied by $\varphi$,
 $|\varphi(s)  -\varphi(t)| = |\varphi(s-t)\varphi(t)   -\varphi(t)| = |\varphi(s-t) -1|
 <\epsilon$, which what we wished to show.

Since $G$ is locally compact, it is complete (see Corollary 1 in
Chapter 3.3, \cite{Bourbaki}). Since $E$ has no interior, $G$ is the
closure  of $D$. Since $\varphi$ is uniformly continuous on $D$ it
extends {\em uniquely} to a uniformly continuous  function $\chi : G
\longrightarrow U(1)$. Then the mapping  $G \times G \longrightarrow
U(1)$ defined by  $(x,y) \mapsto \chi(x+y)\chi(x)^{-1} \chi(y)^{-1}$
is continuous and is equal to $1$ on all of the set $D^{(2)}$. By
Lemma \ref{lemdense} $D^{(2)}$ is dense in $G\times G$ and so, by
the continuity, this mapping is identically equal to $1$. Thus $\chi$ is a
character. \qed

\section{From correlations to model sets} \label{c2ms}

Model sets are an important modeling tool in the subject of quasicrystals.
However, recognizing them is awkward because the cut and project
schemes that underlie them are not obvious from the model sets themselves.

Here we quickly go over a construction given in \cite{BM}, that allows
one to recreate an irredundant cut and project scheme for a model
set $\gL$, using the $2$-point correlation measure $\gamma^{(2)}$.

We know that $\gamma^{(2)}$ is supported within the
uniformly discrete set $\gL-\gL$,
which is a Meyer set. Let $L$ be
the subgroup of $\RR^d$ generated by the set $\gL-\gL$.
For each $\epsilon >0$ define
\[P_\epsilon:=\{t\in L: \dens((t+\gL)\,\triangle \, \gL)<\epsilon\}\,,\]
where $\triangle$ is the symmetric difference operator.
We can use the $P_\epsilon$, $0 \le \epsilon < 2\, \dens(\gL)$, as a
neighbourhood basis of $0$ of a topology, called the
{\em autocorrelation topology}, on $L$ that makes
it into a topological group.
$P_\epsilon$ is called the set of the $\epsilon-${\em almost periods}
of $\gL$.

It is a notable fact about model sets that the sets
$P_\epsilon$ are relatively dense, a consequence of the uniform distribution,
see Thm.~3 \cite{BM}.

Now we define $H$ to be the (Hausdorff) completion of $L$ in the
autocorrelation topology. Then a uniformly continuous homomorphism $\varphi:
L\rightarrow H$ exists with the following properties:
\begin{enumerate}
\item[i)] the image $\varphi(L)$ is dense in $H$;
\item[ii)] the mapping $\varphi$ is an open mapping from $L$ onto
$\varphi(L)$, the latter with the induced topology of the completion
$H$;
\item[iii)] $\mbox{ker}(\varphi)=$ closure of $\{0\}$ in $L$.
\end{enumerate}
Moreover, since $P_\epsilon$ is precompact in the AC topology, $H$
is a locally compact Abelian group.

\begin{prop} $H$ is compactly generated.
\end{prop}

{\bf Proof:}
$L$ is finitely generated (\cite{MoodyNato}, Prop.~7.4), say with finite generating set $J= -J$. Then for any relatively compact open neighbourhood $U$ of $\{0\}$ in $H$,
$V := \bigcup_{j\in J} (U+\varphi(j))$ is also open and relatively compact.
 Let  $V_n:= V+ \cdots + V$ \,  ($n$ summands). Then $\bigcup_{n=1}^\infty  V_n
= H$. In fact, if $x \in H$ then there exists $ y \in (x - U) \,\cap \,\varphi(L)$ since
$\varphi(L)$ is dense in $H$. Writing
$y = \varphi(j_1 + \dots + j_n)$ for some $j_1, \dots, j_n \in J$,
we have $x = u + \varphi(j_1) + \dots + \varphi(j_n) \in V_n$ for some $u \in U$. Thus $\overline V$ is compact and generates $H$. \qed
\smallskip

Now we define
$\tilde{L}:=\{(t,\varphi(t)): t\in L\}$. Then
$\tilde{L}$ is uniformly discrete and relatively dense in $\RR^d\times
H$. Hence, $\tilde{L}$ is a lattice of $\RR^d\times H$ and
$\CalS := (\RR^d, H, \tilde{L})$, along with the natural projection maps, is a cut and project scheme. We introduce the mapping $^\star: L \longrightarrow H$
as above. It is nothing other than $\varphi$.

\begin{theorem}
$\Sigma := \overline {\gL^\star} \subset H$ is compact and is the closure of its interior.
Furthermore, $\gL = \vL(\gO)$ for some $\gO$ lying between
$\Sigma$ and its interior.
\end{theorem}
This result can be found in \cite{BLM} Prop.~4 and Prop.~6.

\smallskip

Although we see now that $\gL$ is a model set from the cut and project
scheme $\CalS$, and we have constructed  $\CalS$ from $\gamma^{(2)}$,
nonetheless, as we pointed out in the Introduction (see \cite{BG}), $\gamma^{(2)}$ does not contain enough information to determine the window.

\section{The role of the $3$-point correlation}

Let $\gL = \vL(\gO)$ where $\gO$ is a window in $H$ with boundary of measure $0$.
Let  $E = E(\gO):=\{k\in \hat H: \widehat{\chF_{\Omega}}(k)=0\}$, and
$D:= \hat H\setminus E$. Since ${\bf 1}_\gO$ is compactly supported and
measurable, it has a continuous Fourier transform $\widehat{\chF_{\Omega}}$. The set $E$ is closed, so $D$ is an open set.  Moreover, since $\widehat{ {\bf 1}_\gO}(0)=\ell(\Omega)>0$, we have $0\in D$.  Measure $0$ alterations to $\gO$ do not affect $E$ and $D$.
The relevance of $D$ and $E$ is immediate from \eqref{diffractionEq}.

\smallskip
We are going to show that the $3$-point correlation is decisive in characterizing a model set
(amongst all other model sets) as long its window $\gO$ in its irredundant
cut and project scheme satisfies $E(\gO)^\circ = \emptyset$.

For real internal groups this is assured:

\begin{prop}If $H = \RR^m$ then the sets $E(\gO)$ have no interior points.
\end{prop}

This is a consequence of the Paley-Wiener theorem \cite{RS}, Thm.~IX.12
since $\Omega$ has compact closure. However, it is also easy to see directly:

{\bf Proof}: Let $g$ be the real part of the function $\widehat{\chF_{\Omega}}$.
For $k \in \RR^m$,
\[g(k)=\int_\Omega \cos(2\pi k\cdot x)dx \, .\]
Writing out the Taylor expansion of $\cos$ around the origin and
integrating term by term, it is easy to see that $g$ has a Taylor
expansion valid on all of $\RR^d$ and hence is an analytic function.
It follows that its zero set in $\RR^d$ has no interior. Since
$\widehat{\chF_{\Omega}}(k)=0$
only if $g(k)=0$, we conclude that the set $E$ has no interior.  \qed

\smallskip
Let us return to the general situation with an internal group $H$.
Let $\gO \subset H$ be
a window in $H$ and assume that $E(\gO)$ has no
 interior points. For notational simplicity let $f:= \chF_{\Omega}$.

Since $D$ is open and $0\in D$, there is a $r_0>0$ such that the cube
$C_{r_0}\subset D$.

We define functions
$\mathcal{I}^n:H^n \longrightarrow \RR$, $n =1,2, \dots $, by
\[
\mathcal{I}^{(n)}(w_1,\dots,w_n):=\ell(\Omega\cap\bigcap_{j=1}^n(w_j+\Omega)),
\]
or equivalently,
\begin{equation}\label{eqdefnofIk}
\mathcal{I}^{(n)}(w_1,\dots,w_n)=\int_{\Rd}\prod_{j=1}^n\tilde{f}(w_j-t)f(t)dt,
\end{equation}
where $\tilde{f}(w):=f(-w)$.

\begin{lemma}
For $n\in\NN$, $\mathcal{I}^{(n)}$ is uniquely determined by
$\gamma_\gL^{(n+1)}$.
\end{lemma}
{\sc Proof:} It is clear that $\mathcal{I}^{(n)}$ is a continuous
function supported inside $(\Omega-\Omega)^n$. Since $\gL^\star$ is
dense in $\Omega$, it follows that $(\gL-\gL)^\star$ is dense in
$\Omega-\Omega$. Moreover, for $x_1,\dots,x_n\in \gL-\gL$,
\[
\mathcal{I}^{(n)}(x_1^\star,\dots,x_n^\star)=
\gamma_\gL^{(n+1)}(-x_1,\dots,-x_n) \, ,
\]
which implies that $\mathcal{I}^{(n)}$ is uniquely determined by
$\gamma_\gL^{(n+1)}$. \qed

\smallskip

Up to density $0$ changes, the compact set $\gO$  is equally well described by its characteristic
function $f = {\bf 1}_\gO$ or by its Fourier transform  $\hat f$. Using the latter
we shall show that $\gO$ is
determined up to translation by $\gamma^{(2)}_{\gL}$
and $\gamma^{(3)}_{\gL}$.

Define $\phi_0$ on $D$ by
\[\phi_0(k):=\frac{\hat{f}(k)}{|\hat{f}(k)|} \, .\]
Then $\phi_0(k)$ is a
continuous function on $D$ and $|\phi_0(k)|\equiv 1$. Since
$\hat{f}(0)=\ell(\Omega)>0$, $\phi_0(0)=1$.
A simple computation shows that
\begin{equation}\label{eqFTI3}
\widehat{\mathcal{I}^{(n)}}(k_1,\dots,k_n)=\prod_{j=1}^n\overline{\hat{f}}(k_j)\hat{f}
(\sum_{j=1}^n k_j) \,.
\end{equation}

When $n=1$
\begin{equation*}\label{eqdefnofI1}
\mathcal{I}^{(1)}(k):=\int_{\Rd}\tilde{f}(k-t)f(t)dt \,.
\end{equation*}
Thus  $\mathcal{I}^{(1)}$
is the convolution product of the function $f$ and $\tilde{f}$ and
\begin{equation*}\label{eqFTI2}
\widehat{\mathcal{I}^{(1)}}(k)=\hat{f}(k)\bar{\hat{f}}(k)=|\hat{f}(k)|^2 \, .
\end{equation*}

When $n=2$,
\begin{equation}\label{eqFTI3}
\widehat{\mathcal{I}^{(2)}}(k_1,k_2)=\overline{\hat{f}}(k_1)\overline{\hat{f}}(k_2)\hat{f}
(k_1+k_2).
\end{equation}

Let $D^{(2)} := \{(k_1,k_2): k_1,k_2,k_1+k_2\in D\}$.
By the definition of $r_0$, $C_{r_0/2}\times
C_{r_0/2} \subset D^{(2)}$. On $D^{(2)}$ we define
\begin{equation*}\label{eqJ3}
\psi^{(2)}(k_1,k_2):=\frac{\widehat{\mathcal{I}^{(2)}}(k_1,k_2)}{|\hat{f}(k_1)||\hat{f}(k_2)||\hat{f}
(k_1+k_2)|}.
\end{equation*}
By (\ref{eqFTI3}), for $(k_1,k_2)\in D^{(2)}$,
\begin{equation}
\phi_0(k_1+k_2)=\phi_0(k_1)\phi_0(k_2)\psi^{(2)}(k_1,k_2).
\end{equation}
This implies that the function $\phi_0$ is a particular solution of
the equation
\begin{equation}\label{eqtheta}
\phi(k_1+k_2)=\phi(k_1)\phi(k_2)\psi^{(2)}(k_1,k_2),
\end{equation}
where  $\phi$ is defined on $D$ and $(k_1,k_2)\in D^{(2)}$. We point
out here that this equation is entirely determined by the functions
$\mathcal{I}^{(1)}, \mathcal{I}^{(2)}$ since the function
$\psi^{(2)}$ is given by them.

Equation (\ref{eqtheta}) is related to the `homogeneous'
equation
\begin{equation}\label{eqthetah}
\varphi(k_1+k_2)=\varphi(k_1)\varphi(k_2), \quad\varphi(0)=1,
\end{equation}
where $\varphi$ is defined on $D$ and $(k_1,k_2)\in D^{(2)}$.

Let $\phi$ be an arbitrary solution of equation (\ref{eqtheta}).
Then $\phi /\phi_0$ is a solution of equation
(\ref{eqthetah}) and by Prop.~\ref{ppcharacter} (with $G$ there
being replaced with $\hat H$) it is
the restriction to $D$ of a character on $\hat H$; that is,
it is the restriction to $\hat H$ of an element of $\hat{\hat H} \simeq H$. Hence
it is determined by some element $a \in H$ through $ \chi_a: k\mapsto \langle k,a\rangle \in U(1)$ and each solution of (\ref{eqtheta}) has
the form
\begin{equation}\label{eqphi}
\phi(k):=\phi_0 (k)\chi_a(k) \,.
\end{equation}

Finally, we get the main result of this section.
\begin{theorem}\label{main}
Let $\gL=\vL(\Omega)$ be a regular model set and assume
that $E(\gO)$ has no interior points (which is
guaranteed if $H$ is a real group). Then any
model set with the same  2-point and 3-point
correlation measures as $\gL$ is, up to density $0$,
a translate of $\gL$.
\end{theorem}
\noindent {\bf Note:} All elements of the hull differ from translates of $\gL$ by
sets of density $0$, see \cite{BLM}, Prop.~7.
\smallskip

{\sc Proof:} Let $\gL'$ be a model set for some irredundant cut and project  scheme
$\CalS'$ and suppose that it has the same $2$- and $3$-point correlations
as the model set $\gL = \gL(\gO)$ from the irredundant cut and project
scheme $\CalS$. The equality of the $2$-point correlations shows that
we can assume that $\CalS' = \CalS$.

Recall that equation (\ref{eqtheta}) is determined by
$\widehat{\mathcal{I}^{(1)}}$ and $\widehat{\mathcal{I}^{(2)}}$.
Since $\gL'$ has the same $2$- and $3$-point correlations as $\gL$,
the function $\phi$ for $\gL'$ corresponding
to the function $\phi_0$ for $\gL$ is another solution of \eqref{eqtheta}.
As we have already shown, this implies that
\eqref{eqphi} holds
for some  $a\in H$.

Let $f'$ be the characteristic function of the window of $\gL'$. Then
$\phi = \widehat{f'} / |\widehat{f'}|$
and
$|\widehat{f'}|=(\widehat{\mathcal{I}^{(1)}})^{\frac{1}{2}}=|\hat{f}|$.
Thus
\[
\widehat{f'}(k):=\hat{f}(k)\chi_a(k).
\]
Taking the inverse Fourier transform on both sides of this equation
we have
\[
f' \sim \chF_{-a+\Omega} \, ,
\]
i.e. they are equal except possibly on a set of measure $0$, which shows that up to density $0$, $\gL'$ is
a translate of $\gL$. \qed

\section{Counterexamples}\label{CE}

In this section we give examples of cut and project schemes and model sets
from them for which the $2$- and $3$-correlations do
not determine the window of the model set up to translation or inversion.
For these sets the corresponding sets $E(\gO)$ have interior points.

\subsection{A periodic example}\label{PE}
The most obvious way to make the sets $E(\gO)$ have interior points is to use a compact internal space, for
then $\hat H$ is a discrete group. In the case that $H$ is a finite group we are dealing with periodic structures, and for these it is long known that $2$- and $3$-correlations may fail to characterize the structure. The paper of Gr\"unbaum and Moore \cite{GM} has useful introduction
to the homometry problem for crystals (periodic structures) and some
good ways of making examples. One such example, put here into the language
of model sets, is the following:

With $N = 32$,  there are sets $A,B \subset \{0, \dots, 31\}$
which have the same $2$- and $3$-point correlations (when the point patterns
are treated modulo $N$) \cite{GM} \S5.3,
but which are not equivalent by `rigid motions'.  That is, they have the same
pattern frequencies in $\ZZ/N \ZZ$ for all $2$- and $3$-point patterns.
Explicitly such a pair is given by the exponents of the expansions of the polynomials
\begin{eqnarray*}
p_A(x) := \frac{1 - x^{16}}{1-x} (1 - x^3 + x^9)(1 - x + x^3 - x^4 +x^6)
=1 + x^7 + x^8 + x^9 + x^{12}\\ + x^{15}
 + x^{17} + x^{18}
 + x^{19} + x^{20} + x^{21} + x^{22}
 + x^{26} + x^{27} + x^{29} + x^{30}
 \end{eqnarray*}
 \begin{eqnarray*}
p_B(x):=\frac{1 - x^{16}}{1-x} (1 - x^3 + x^9)(1 - x^2 + x^3 - x^5 +x^6)
=1 + x + x^8 + x^9 + x^{10} \\+ x^{12}+ x^{13} + x^{15} + x^{18} + x^{19} + x^{20} +
x^{21} + x^{22} + x^{23} + x^{27} + x^{30} \, .
\end{eqnarray*}

That is, $A = \{0,7,8, 9,12, \dots, 30\}$ and $B = \{0,1,8, 9,10, \dots, 30\}$.
Comparing the gaps between successive positions in $A$ and $B$
along the line, namely ($6, 0,0,2, \dots, 1,0;1)$ and $(0,6,0,0, \dots. 3,2;1)$ where
we wrap around modulo $32$ at the semicolon, one sees directly that the gap
patterns are neither translationally equivalent or equivalent by reversal of
direction.

Now form the cut and project scheme

\begin{equation} \label{cpFiniteScheme}
   \begin{array}{ccccc}
      \RR & \stackrel{\pi_{1}}{\longleftarrow} &
        \RR \times \ZZ/N\ZZ & \stackrel{\pi_{2}}
      {\longrightarrow} & \ZZ/N\ZZ   \\
      &&  \cup \\
      L&\stackrel{\simeq}{\longleftrightarrow} & \widetilde \ZZ &\qquad &\qquad\\
      x & \longleftrightarrow& (x,x_N) & \mapsto & x_N \,,
   \end{array}
\end{equation}
where $x_N:= x \mod N$.

Then for $A$ and $B$, taken modulo $N$, we have $\vL(A), \vL(B)$. These are periodic
subsets of $\ZZ$ which are located at the points of $A +N \ZZ$ and
$B+N\ZZ$. To keep things straight below, we will denote them
by $\vL_p(A), \vL_p(B)$, indicating their periodic nature. The pattern frequencies for $2$- and $3$-point patterns of $\vL_p(A)$ and $\vL_p(B)$ are not altered when reduced$\mod N$, and hence are equal.  Assuming normalization so that
the density of $\ZZ$ is $1$, the appropriate volume function on $\ZZ/N\ZZ$
is $\vol_{\ZZ/N\ZZ}(S) = \card(S)/N$ for all $S \subset \ZZ/N\ZZ$.

This produces two (periodic) model sets which are not related by translation
or inversion of windows but which have the same $2$- and $3$-point correlations.
The diffraction from either of the two model sets is the same. It is periodic of period $1$
and supported on the set $\frac{1}{32}\ZZ$ with values explicitly given by
\[ \left| \frac{1}{32} p_A(\exp(2 \pi i k/32)\right|^2 \delta_{\frac{1}{32}\ZZ}(k)  \,\]
at $k/32$, $k \in\ZZ$. (The label $A$ can be replaced by $B$ here).

\begin{figure}
\centering
\includegraphics[width=12cm]{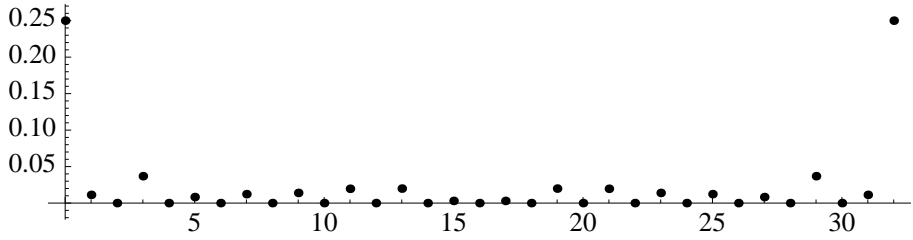}
\caption{A plot over a complete cycle of the diffraction of the periodic model set determined by the set $A$ of \S7.1. The tick mark labels $k$ stand for
$k/32$ and the domain is shown for $k=0, \dots, 32$. Note the vanishing of the diffraction
(extinctions) at all points $2k/32$, $k \not\equiv 0 \mod 16$. }
\label{periodicDiffraction}
\end{figure}

\subsection{An aperiodic example}\label{Aperiodic}
We can use this example to create distinct aperiodic model sets on the line which
have the same $2$- and $3$-point correlations. The example below, based on the
standard Fibonacci model sets, shows how this can be done.

Let $\tau := (1+ \sqrt 5)/2$, let $L := \ZZ + \ZZ \tau$, and let
$':L \longrightarrow L$ be the conjugation mapping defined
by $\tau' := (1- \sqrt 5)/2$ . Form the standard Fibonacci cut and project scheme
\begin{equation} \label{FibcpScheme}
   \begin{array}{ccccc}
      \RR & \stackrel{\pi_{1}}{\longleftarrow} &
        \RR \times \RR & \stackrel{\pi_{2}}
      {\longrightarrow} & \RR   \\
      &&  \cup \\
      L&\stackrel{\simeq}{\longleftrightarrow} & \cL &\longrightarrow & L \\
      x & \longleftrightarrow& (x,x') & \mapsto & x'\,.
   \end{array}
\end{equation}
Each $x = u  + v \tau \in L$, $ u,v \in \ZZ$, is mapped to $(x,x') = u(1,1) + v(\tau,\tau')
\in \cL$. We
define $\alpha: L\longrightarrow \ZZ/N\ZZ$ by $\alpha(x) =  u_N$, i.e.
reduction of $u$ modulo $N$.  Let $W$
be a window in $\RR$ and $\vL(W)$ the corresponding model set.

Now consider the combined cut and project scheme:

 \begin{equation} \label{BigcpScheme}
   \begin{array}{ccccc}
      \RR & \stackrel{\pi_{1}}{\longleftarrow} &
        \RR \times (\RR \times \ZZ/N\ZZ) & \stackrel{\pi_{2}}
      {\longrightarrow} & \RR \times \ZZ/N\ZZ    \\
      &&  \cup \\
      L&\stackrel{\simeq}{\longleftrightarrow} & \cL_c &\longrightarrow &
      L \times \ZZ/N\ZZ  \\
      x & \longleftrightarrow& (x,(x', \alpha(x))) & \mapsto & x^\star :=(x', \alpha(x))\,.
   \end{array}
\end{equation}
Notice here that $\cL_c :=\{(x,(x', \alpha(x))) \,:\, x\in L\}$.
With the subsets $A, B$ above, we obtain the model sets $\vL_c(W\times A)$
and $\vL_c(W \times B)$, where the subscripts $c$ stand for the
{\em combined} cut and project scheme. In effect these consist of the points $x$ of $\vL(W)$ for which $\alpha(x) \in A$ (resp. $B$).

\begin{figure}
\centering
\includegraphics[width=13cm]{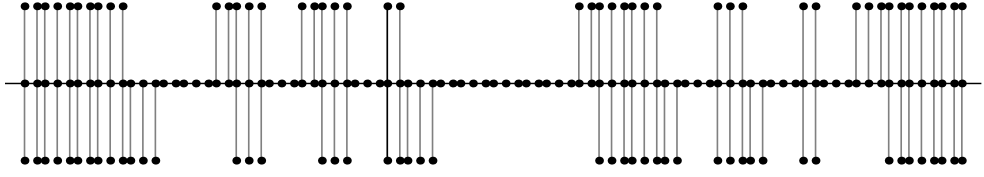}
\caption{A fragment the Fibonacci model set $\{ \dots, -1-\tau, -1,0, \tau, 1+\tau, \dots \}$ based on the window $[-1,\tau^{-1})$
and the correspondingly thinned model sets using the congruence sets $A$ and
$B$ of \S7.1 above and below it respectively. These two thinned sets have the same
second and third moments, although they are not translationally equivalent. The location of the origin is indicated by the black vertical line.}
\label{thinnedModelsets}
\end{figure}

Now consider any pattern $\{0,x,y\} \subset L$ for $\vL_c(W\times A)$.
Write $x^\star = (x',r)$, $y^\star = (y' ,s)$ with $r,s \in A$. Up to the appropriate normalizing factor (see Thm.~\ref{uniformDistribution}), the frequency of the
pattern in $\vL_c(W\times A)$, which is also the $3$-point correlation $\gamma^{(3)}((x,y))$, is
\begin{eqnarray*}
\frq(\{0,x,y\}) &=& \vol_c( (W \times A) \cap(-x^\star + (W\times A))
\cap (-y^\star +(W\times A))\\
&=&\vol_c( (W  \cap(-x' + W) \cap (-y' +W)
\times  \\
&\quad& \qquad \qquad \qquad \qquad A \cap(-r + A)) \cap (-s +A))\\
&=& \vol_\RR( (W  \cap(-x' + W) \cap (-y' +W))\, .\\
&\quad& \qquad \qquad \qquad   \vol_{\ZZ/N\ZZ}(A \cap(-r + A)) \cap (-s +A)) \, .
\end{eqnarray*}
Replacing $A$ by $B$ gives the three point correlation at $(x,y)$ for
$\vL_c(W\times B)$. However
\[\vol_{\ZZ/N\ZZ}(A \cap(-r + A) \cap (-s +A)) = \vol_{\ZZ/N\ZZ}(B \cap(-r + B) \cap (-s +B))\]
and so we have shown that $\vL_c(W\times A)$ and $\vL_c(W\times B)$
have the same $3$-point correlations (and hence also equal $2$-point correlations).
Due to the different gap patterns produced by $A$ and $B$ the windows
of the two model sets cannot be matched by translation or inversion.

\subsection{Aperiodic $\times$ periodic} Even simpler is to construct model sets of the form aperiodic $\times$ periodic
with identical $2$- and $3$-point correlations
Suppose that $\Sigma := \vL(W)$ is a model set arising from the cut and project
scheme \eqref{cpScheme}. We take the direct product of this cut and project
scheme and the one given at \eqref{cpFiniteScheme}, and take as the windows
the sets $W \times A$ and $W \times B$. This leads to model sets
$\Sigma \times \vL(A)$ and $\Sigma \times \vL(B)$.

Each instance $\{z, z+z_1, z+z_2\}$ in $\Sigma \times \vL(A)$ of a pattern $\{0,z_1, z_2\}$ in $\RR^d \times
\ZZ/N\ZZ$ is uniquely expressible as instances
\[\{x, x+x_1, x+x_2\} \times \{a, a+a_1, a+a_2\}\]
from  $\Sigma$ and $\vL(A)$
of the product
\[\{0, x_1, x_2\} \times \{0, a_1, a_2\}\]
of patterns from $\RR^d$ and $\ZZ/N\ZZ$; and vice-versa.
The frequency of $\{0,z_1, z_2\}$ in $\RR^d \times \ZZ/N\ZZ$ is the product
of the frequencies of $\{0, x_1, x_2\} $ in $\Sigma$ and $\{0, a_1, a_2\}$
in $\vL(A)$.

The same goes when we use $\vL(B)$, and so $\Sigma \times \vL(A)$ and $\Sigma \times \vL(B)$ have the same $2$- and $3$-point correlations.

\subsection{Comments on real spaces $\times$ finite groups}

Let $F$ be a finite group (with the discrete topology) and $\hat F$ its dual (also finite). We are interested in
the Fourier transforms of characteristic functions on subsets of
spaces of the form $\RR^m \times F$, and in conditions under which they may
have zero sets with non-empty interiors.

Give $\RR^m \times F$ the product topology. Let $\gO \subset \RR^m \times F$ be a non-empty relatively compact
set satisfying $\overline{\gO} = \overline{\gO^\circ}$ and write
\[ \gO = \bigcup_{a\in F} (\gO_a, a) \, .\]
Let $A$ be the set of elements $a  \in F$ for which $\gO_a$ has a non-empty
interior. Then $\widehat{\chF_\gO}: \widehat{\RR^m} \times \hat F \longrightarrow
\CC$:
\[\widehat{\chF_\gO}(k,b) = \sum_{a\in A} \int_{\gO_a} e^{-2\pi i k\cdot x} \overline{\chi_a(b)}\, dx
= \sum_{a\in A}  \overline{\chi_a(b)}\,\widehat {\chF_{\gO_a}}(k) \]
where $\chi_a$ is the character on $\hat F$ defined by $a\in F$.

For each fixed $b \in \hat F$, this is an analytic function of $k$ on $\widehat{\RR^m}
\simeq \RR^m$.
Let $E_b$ denote the vanishing set of $\widehat {\chF_\gO}$ on
$\widehat{\RR^m} \times \{b\}$. In order that $\widehat {\chF_\gO}$ vanish
on a set with interior, we require that some of the $E_b$ have interior points.
Being analytic on $\RR^m \times \{b\}$, this requires
\[\sum_{a\in A}  \overline{\chi_a(b)}\,\widehat {\chF_{\gO_a}}(k) \equiv 0 \,\]
and hence that
\begin{equation} \label{zeroCondition}
\sum_{a\in A}  \overline{\chi_a(b)}\,\chF_{\gO_a}(k) \equiv 0 \,
\end{equation}
as a function on $\RR^d$.

There are straightforward ways to make this happen. For instance,
if all $\gO_a$, $a\in A$, are equal then the requirement is simply
that $\sum_{a\in A} \chi_a(b) = 0$ for some $b\in \hat F$, which is certainly possible.

This is what is going on in the examples above. The equation
$\sum_{a\in A} \chi_a(b) = 0$ becomes
\[p_A(e^{2 \pi i b/N}) = \sum_{a\in A} e^{2 \pi i a\,b/N} = 0\,,\]
where $b \in \{0,1, \dots 31\}$.
This equation holds whenever $b$ is even and different from $0$, since then the factor $1 + x + \dots + x^{15}$ of $p_A(x)$ evaluates to $0$. Similarly for $B$.

On the other hand, for the Penrose
point sets arising from the Penrose rhombic tiling, the vanishing set $E$
is indeed without any interior points. In this case the internal space is
$\CC \times \ZZ/5\ZZ$ and the windows are the pentagons
$P, -\tau P, \tau P, -P$ where $P$ is the convex hull of the $5$th roots of $1$
and the four listed windows are for congruence classes $1,2,3,4 \mod 5$
respectively \cite{MoodyNato}. Here it is easy to see that the
zero condition \eqref{zeroCondition} cannot possibly be satisfied.
So rhombic Penrose point sets {\em are} determined
by their $2$- and $3$-point correlations.

One final comment. One can raise the bar and ask about examples with
equal $4$-point or higher correlations. There are no examples of the type that we
have produced here, because there are no finite cyclotomic sets
like $A, B$ above which are translationally inequivalent but have equal
$4$-point correlations, \cite{GM}. However, extending the investigation to
{\em weighted} point sets, there are examples with equal second through
fifth point correlations, \cite{DM2}.

\section{Connections with point processes} \label{cpp}

Start with the cut and project scheme \eqref{cpScheme}. Let $\gO \subset H$ be a window (note the assumptions on windows, given in \S\ref{ms}) for a model set $\gL = \gL(\gO)$ in $\RR^d$. Let $\Sigma := \overline{\gO}$.
Then $\gL$ is a Delone set and there is a positive number $r$ so that
its distinct points are separated by at least the distance $r$.
Let $\CalD_r$ be the family of all
discrete subsets of $\RR^d$ whose points are separated by at least the
distance $r$. There is a uniformity on $\CalD_r$ whose entourages are
generated by the sets
\[ U_{R,\epsilon} := \{ ( \Phi, \Phi') \in \CalD_r^2 \, :\,
\Phi \cap C_R \subset \Phi' + C_\epsilon \; \mbox{and} \;
\Phi' \cap C_R \subset \Phi + C_\epsilon \} \]
as $R, \epsilon$ vary over $\RR_{>0}$.

The topology defined by this uniformity is a Fell topology, and so $\CalD_r$ is a compact
Hausdorff space (\cite{Fell}, Thm.1).
The translation action of $\RR^d$ on $\CalD_r$ is continuous and the
pair $(\CalD_r, \RR^d)$ is a topological dynamical system \cite{Radin}.

The {\em hull} $X(\gL)$ of $\gL$ is the closure in this topology of the set of
all translates of $\gL$. It is clearly a compact subset of $\CalD_r$ and is also translation invariant. Thus $(X(\gL), \RR^d)$ is a dynamical system in its own right.

It is known \cite{Schlottmann} that $X(\gL)$ is uniquely ergodic. Let
$\mu$ denote the unique ergodic probability measure on it.
There is a continuous $\RR^d$-equivariant mapping $\beta: X(\gL) \longrightarrow
\TT$, called the {\em torus parametrization}, from the hull into the group
$\TT$ of the cut and project scheme \eqref{cpScheme}. The inverse image
in $X(\gL)$ of any point $\xi = (x,y) \mod \cL$ is made up of sets of the form
$ x + \vL(-y + \gO')$
for some set $\gO'$ satisfying $\Sigma^\circ \subset \gO' \subset \Sigma$
\cite {Schlottmann, BLM}. If $\gL$ is a regular model set, then so are
all the elements of $X(\gL)$. Furthermore, assuming regularity, $\mu$-almost surely the inverse image of $\xi = (x,y) \mod \cL$ consists of just one point and it is
$x + \vL(-y + \Sigma^\circ) = x + \vL(-y + \Sigma)$. These elements are called
{\em non-singular} elements of $X(\gL)$. See \cite{BLM}, \S 5.3 for more
details.

Thus the point sets that make up $X(\gL)$ are model sets and almost all
of them are {\em non-singular}.
All of the point sets in $X(\gL)$ have point correlations of all orders $k$, and these
$k$-point correlations are identical for all the elements of $X(\gL)$. Thus we may
speak of the $k$-point correlation of $X(\gL)$.

There is an obvious mapping $N$ from the bounded measurable
subsets of $\RR^d$ into $L^2(X(\gL), \mu)$ defined by
\[ N(A)(\Phi) = \card(\Phi \cap A) \,.\]
The mapping $N$ can be construed as an ergodic uniformly discrete point process, with
$\int_{X(\gL)} N(A) d \mu $ being the expectation for the number
of points in $A$ for a randomly chosen (according to the law $\mu$)
point set $\Phi \in X(\gL)$. The function $N$ extends naturally to a
 $\RR^d$-equivariant mapping $N: C_c(\RR^d) \longrightarrow L^2(X(\gL),\mu))$ defined by
\[ N(f)(\Phi) = \sum_{x\in \Phi} f(x)  \,\]
for all $\Phi \in X(\gL)$.

Following \cite{Gouere}, there has been increasing
interest in studying the mathematics of quasicyrstals by using ideas from
the theory of point processes.
Although quasicrystals (and model sets!) are, by their very nature, assumed to
be highly ordered (correlated),
and hence quite atypical from the point of view
of the theory of point processes, nonetheless the techniques of
this theory are applicable and quite effective. For a more detailed and
comprehensive exposition of this point of view
see \cite{DM}.

There is a one-one correspondence between the correlation measures of a typical
point set and the moment measures of the corresponding point processes \cite{DVJ},
\S12.2.  The correspondence between the 2-point correlation
and the first reduced moment measure was utilized by
Gou\'{e}r\'{e} \cite{Gouere} in his analysis of pure point diffraction
and almost periodicity. In \cite{DM}, we prove that an ergodic
point process of uniformly discrete point sets is uniquely
determined by its moments, or equivalently, all of its correlation measures.

This implies that
solving the inverse problem for quasicrystals (i.e. determining the law
of the corresponding point process) is equivalent to knowing
all of its $k$-point correlations. In this language we have from Thm.~\ref{main}:

\begin{theorem} \label{mainPointProcess} Let $X(\gL)$ and $X(\gL')$ be point processes from
regular model sets and suppose that their second and third moments are
the same. Then their cut and project schemes may be identified. If
$\TT$ is the corresponding compact Abelian group and $\beta, \beta'$ are
corresponding torus mappings, then for each $\xi \in \TT$ the
elements of $\beta^{-1}(\xi)$ and $\beta'^{-1}(\xi)$ all differ from one
another at most on sets of density $0$.
 \end{theorem}

Another approach to the way in which correlations link to the structure of the dynamics in the pure point case is given in \cite{LenzM}. Here the setting is
a pure point ergodic uniformly discrete point process $(X,\mu,N)$. The diffraction is almost always the same for the $\Lambda \in X$ and it the Fourier transform of the first moment of the Palm measure of the process. The group $M$ generated by the set $S = -S$ of positions of the Bragg peaks is the Fourier module of the process.  There it is proved that if
$S + \dots +S = M$, where there are $n$ summands, then the point process is
determined by the $2$-, $3$-, \dots, $(2n+1)$-point correlations. Thus if there
are no extinctions ($S=M$) we need only the $2$- and $3$-point correlations.
This result is not as strong as we have obtained here for real model sets with real internal spaces, since we have made no requirements on $S$.
On the other hand, it goes well beyond model sets and works for general locally compact Abelian groups, and furthermore suggests a fundamental role for the extinctions in understanding how much of the dynamics is controlled by the diffraction.

\section{Acknowledgments}

The authors are very grateful to Gabriele Bianchi for his interest in this
paper and for pointing us to some of the relevant literature on higher correlations
which became the basis of the counterexamples here; to Ron Lifshitz for helping us understand the underlying philosophy of D.~Mermin's paper \cite{Mermin};
and to Michael Baake for his continuing interest and support. Thanks also to the referees of this paper for their helpful suggestions which have improved the paper.

RVM acknowledges the support of the Natural and Engineering Research
Council of Canada.

\end{document}